# Influential Mathematicians: Birth, Education and Affiliation

J. Panaretos[1] and C.C. Malesios
Department of Statistics
Athens University of Economics and Business
76 Patision St, 10434 Athens Greece

*Abstract*—Research output and impact is currently the focus of serious debate worldwide. Quantitative analyses based on a wide spectrum of indices indicate a clear advantage of US institutions as compared to institutions in Europe and the rest of the world. However the measures used to quantify research performance are mostly static: Even though research output is the result of a process that extends in time as well as in space, indices often only take into account the current affiliation when assigning influential research to institutions. In this paper, we focus on the field of mathematics and investigate whether the image that emerges from static indices persists when bringing in more dynamic information, through the study of the "trajectories" of highly cited mathematicians: birthplace, country of first degree, country of PhD and current affiliation. While the dominance of the US remains apparent, some interesting patterns -that perhaps explain this dominance- emerge.

## I. INTRODUCTION

There is currently a surge of interest in comparing research impact and performance, to produce league tables. These may be at various levels, ranking countries, universities, departments, programs, journals or even individual scientists, and are typically based on certain simple bibliometric measures, such as impact factors, the h-index etc.

This interest is not purely academic: these rankings have caught the attention of policy makers, and have caused serious concern especially within European policy making due to the apparent lagging performance of Europe as compared to the US. This has been documented by several indicators and reports commissioned by EU (see, e.g. Saisana and d'Hombres, 2008; Lambert and Butler, 2006; Moed, 2006), but perhaps is best exemplified by the French president's public setting in January, 2008 as an aim to ameliorate the position of French universities in the international rankings. If rankings can affect educational policy at such a high level, it is natural to revisit the question of how accurately they represent the truth, research quality being so difficult to quantify – which is especially true in the field of mathematics.

Criticisms focus on the appropriateness of different measures, their sensitivity/robustness and their interpretability (see, e.g., Adler et al., 2008; Saisana and d'Hombres, 2008; Evidence Report, 2007). For a detailed critical review of such indices, see Panaretos & Malesios (2008).

A different aspect that has not received attention is the static character of several of the indices employed, which fails to capture the "liquidity" of the modern academic landscape, where high mobility of scientists is the rule rather than the exception. This is manifested as a sort of Markovian property: the past is irrelevant given the present. But aside from the most recent affiliation of the scientists considered, is it reasonable to forfeit the movement of scientists at various stages of their career?

To take an example from the field of mathematics: how should the credit of the achievements of Jong-Shi Pang, a highly cited mathematician, (http://www.iese.uiuc.edu/research/faculty/pang.html) be attributed to a country/institution? Jong-Shi Pang was born in Vietnam, obtained his first degree at the National University of Taiwan, completed his PhD at Stanford University, and has been affiliated with the University of Texas at Dallas, Carnegie Mellon University, the University of Wisconsin – Madison, Johns Hopkins University, the Rensselaer Polytechnic Institute, before moving to the University of Illinois at Urbana-Champaign in 2007. While his present affiliation obviously deserves a lot of the credit stemming from his high citations, should we not take into account the fact that the scientist has been "nurtured" and "grown scientifically" in many places?

---

[1] e-mail for correspondence: jpan@aueb.gr

The purpose of this article is to attempt a first probe of the "movement effect" and see how this might influence a concrete question, such as the comparison between the US and Europe in the field of mathematics. We focus on highly cited mathematicians, since citations are often taken as a strong indicator of research impact, and track their countries of birth, education, and current affiliation.

In general, comparable data on researchers' movement between Europe, Asia or Africa to the US are incomplete. A database on highly cited researchers (HCRs) is compiled by the Institute of Scientific Information (ISI) covering 21 disciplines and 6.103 researchers[2]. These data are freely available by the Thomson Scientific (http://hcr3.isiknowledge.com/) and cover the time period between 1981 and 1999.

With regards to mathematics, the Thomson database lists 343 highly cited mathematicians from 152 Institutions. While the Thomson database may provide the list of HCRs and their present affiliation, we had to conduct a personalized case-by-case search in order to obtain data on the country they obtained their first degree, and their PhD as well as their birthplace, either by searching through their webpages or by contacting them directly.

Table A3 summarises the data on HCRs in the field of Mathematics according to the country of their present affiliation. One easily sees that the US – as in all disciplines – gets the lion's share of HCRs. The UK and France are far behind the US, but well ahead of the rest of the countries.

By bringing in the additional background data, we can immediately observe that intercontinental movement is indeed a very common practice. Specifically, based on the data collected, only the 46.9% of HCRs were born, educated and are working in the same continent, while a significant 42.6% of them have completed at least one of their degrees or are working in a continent other than the one they were born in (due to missing information we cannot answer this question for the 10.5% of HCRs). Our findings are presented in more detail in the following sections.

## II. THE EDUCATIONAL BACKGROUND OF HCRS IN THE FIELD OF MATHEMATICS

In this section, we examine the geographical breakdown of the numbers of HCRs in the field of mathematics taking into consideration the country of their birth, the country where their first degree and the country where their PhD degrees were obtained.

### A. Current affiliation of HCRs

Table 1 presents the percentages of HCRs in the field of mathematics according to their current affiliation. The majority of researchers are working in the US (68.2%), while 22.7% work in Europe[3]. Only 9% work in countries outside the US and Europe. (Countries with more than one HCR outside the US and Europe are Israel, Canada, Japan, and China). The percentages in the mathematics discipline are quite analogous to the percentages of all 21 disciplines (see Table A2).

*Table 1: Frequencies and percentages of HCRs according to the country of their present affiliation*

| | | FREQ | (%) |
|---|---|---|---|
| Valid | **US** | 234 | 68.2 |
| | **Europe** | 78 | 22.7 |
| | **Israel** | 8 | 2.3 |
| | **Australia** | 6 | 1.7 |
| | **Canada** | 6 | 1.7 |
| | **Japan** | 5 | 1.5 |
| | **China/Taiwan** | 3 | 0.9 |
| | **India** | 1 | 0.3 |
| | **Singapore** | 1 | 0.3 |
| | **Turkey** | 1 | 0.3 |
| | **TOTAL** | 343 | 100.0 |

Evidently, when looking only at current affiliation, the US dominates most emphatically Europe, which in turn is well ahead of the rest of the world. Will this pattern persist when bringing in more background information?

### B. PhD studies of HCRs

When focusing on the country where HCRs completed their PhD education, the US maintains an advantage over Europe and the rest of the world but not nearly as strong as when compared with respect to current affiliation of the HCRs (Table 2). In particular, 57.7% of HCRs in mathematics have acquired their Ph.D. degree in US universities, 32.1% in Europe and 8.5% in the rest of the world: the difference between the US and Europe drops by approximately 20 percentage points.

---

[2] Table A1 in the appendix provides information on the numbers of HCRs according to the country of their present affiliation. A further break down by scientific discipline of the numbers of HCRs according to the country of present affiliation (US, Europe and the rest of the world) is given in Table A4. As one can observe, US Institutes dominate the list – in terms of HCRs – in the fields of Social Sciences (93.1%), Economics (86.2%), Psychology-Psychiatry (86.1%), Clinical Medicine (75.8%) and Computer Science (73.9%). On the other hand, European institutions have the highest concentration of HSC in the field of Pharmacology (46.8%). In fact, this is the only instance where Europe outperforms the US in terms of HCRs (123 HCRs in comparison to 94 HCRs working in the US). The highest percentage of HCRs working in non-US and EU countries is observed in the Agricultural Sciences field (26.2%).

[3] The majority of European Institutions with HCRs are based in EU countries. Three HCRs are working in Switzerland. In some places we use the term EU with this in mind

*Table 2: Frequencies and percentages of HCRs according to the country where the Ph.D. studies were completed*

|  |  | FREQ | (%) |
|---|---|---|---|
| Valid | US | 198 | 57.7 |
|  | Europe | 110 | 32.1 |
|  | Israel | 7 | 2.0 |
|  | Canada | 6 | 1.7 |
|  | Russia | 5 | 1.5 |
|  | Japan | 5 | 1.5 |
|  | India | 2 | 0.6 |
|  | Australia | 2 | 0.6 |
|  | Argentina | 1 | 0.3 |
|  | South Africa | 1 | 0.3 |
|  | Total | 337 | 98.3 |
|  | Missing | 6 | 1.7 |
|  | TOTAL | 343 | 100.0 |

*Table 3: Contingency table between the country of present affiliation of the HCRs and the country of the Ph.D. degree of the HCRs*

|  |  |  | Country of Present Affiliation of the HCRs |  |  | TOTAL |
|---|---|---|---|---|---|---|
|  |  |  | US | EU | Rest of the world |  |
| Country in which the Ph.D. Degree of the HCRs was obtained | US | Count | 180 | 6 | 12 | 198 |
|  |  | % within | 90.9% | 3.0% | 6.1% | 100.0% |
|  | EU | Count | 37 | 65 | 8 | 110 |
|  |  | % within | 33.6% | 59.1% | 7.3% | 100.0% |
|  | Rest of the world | Count | 16 | 2 | 11 | 29 |
|  |  | % within | 55.2% | 6.9% | 37.9% | 100.0% |
| TOTAL |  | Count | 233 | 73 | 31 | 337 |
|  |  | % within | 69.1% | 21.7% | 9.2% | 100.0% |

The distribution provided in Table 3 reveals that a stunning one in three HCRs who completed their doctorate in Europe is now affiliated with a US institution. Even more extreme is the situation when looking at HCRs with PhDs from outside the US or Europe, one in two of whom have eventually settled in the US.

The above findings outline an overflow of outstanding mathematicians to the US (a phenomenon known as "the brain drain"), which is confirmed to be a significant factor contributing to the global dominance of US Institutions.

The opposite type of movement is very rare, since only 3% and 6.1% of those who have completed their Ph.D. studies in the US have moved to Europe and to non-European countries, respectively. In particular, the percentage of "EU doctors" moving to the US is over ten times higher than the percentage of "US doctors" moving to Europe: it seems that Europe is failing not only to retain their top talent, but is also failing to attract top talent (a more detailed contingency table (A6) is presented in the Appendix).

### C. BSc studies of HCRs

Examination of the country where the HCRs in mathematics earned their first degree reveals further interesting facts (Table 4). Only 32.7% of the HCRs completed their B.Sc. degree studies in the US, while 33.2% completed their first degree in Europe and a quite significant number (25.4%) have completed their B.Sc. studies in countries outside the US and Europe. The distribution of HCRs between the three different "regions" seems close to uniform at this stage. As we go further back into the background of the HCRs, the distribution of HCRs among countries becomes more and more diffuse.

This could be an indication that "promising" undergraduate mathematics students are found equally in Europe and in the US and also in other countries outside the US and Europe.

*Table 4: Frequencies and percentages of HCRs according to the country where the first degree was completed*

|  |  | FREQ | (%) |
|---|---|---|---|
| Valid | EU | 114 | 33.2 |
|  | US | 112 | 32.7 |
|  | China/Taiwan | 18 | 5.2 |
|  | Canada | 14 | 4.1 |
|  | Australia | 11 | 3.2 |
|  | India | 9 | 2.6 |
|  | Russia | 7 | 2.0 |
|  | Israel | 6 | 1.7 |
|  | Hong Kong | 4 | 1.2 |
|  | Japan | 4 | 1.2 |
|  | South Africa | 4 | 1.2 |
|  | rest of the world (*) | 10 | 2.9 |
|  | Total | 313 | 91.3 |
|  | Missing | 30 | 8.7 |
|  | TOTAL | 343 | 100.0 |

*(*) 1 HCR for each of Argentina, Peru, Egypt, Brazil, Mexico, New Zeeland, Venezuela, Algeria, Turkey and Chile*

Table 5 provides a contingency table between the country in which the first degree was completed and the country of present affiliation and allows for more detailed comparisons.

*Table 5: Contingency table between the country of present affiliation of the HCRs and the country where the first degree of the HCRs was completed*

|  |  |  | Country in which the B.Sc. Degree of the HCRs was obtained | | | TOTAL |
|---|---|---|---|---|---|---|
|  |  |  | US | EU | Rest of the world |  |
| Country of Present Affiliation of the HCRs | US | Count | 107 | 50 | 61 | 218 |
|  |  | % within | 49.1% | 22.9% | 28.0% | 100.0% |
|  | EU | Count | 3 | 62 | 2 | 67 |
|  |  | % within | 4.5% | 92.5% | 3.0% | 100.0% |
|  | Rest of the world | Count | 2 | 2 | 24 | 28 |
|  |  | % within | 7.1% | 7.1% | 85.7% | 100.0% |
| TOTAL |  | Count | 112 | 114 | 87 | 313 |
|  |  | % within | 35.8% | 36.4% | 27.8% | 100.0% |

The results indicate a significant transfer of mathematics researchers to the US from the rest of the world, when the first degree is taken into account (from a total of 218 HCRs affiliated with US Institutions, 50 and 61, respectively, have acquired their first degree in Europe and the rest of the world). Notice how diffuse the distribution of HCRs affiliated with US institutions is with respect to the country of their alma mater: only one in two were undergraduates in US universities; the contrast with Europe is stark, as its respective distribution is acutely concentrated: nine out of ten HCRs affiliated with European Institutions also received their bachelor degrees from within Europe.

A more detailed version of the contingency table is presented in the Appendix (Table A5). The majority of highly cited researchers affiliated with US Institutions with B.Sc. studies outside the US and Europe are coming from China, Canada and India (16, 11 and 7, respectively). On the other hand, only 5 HCRs are affiliated with European Institutions having acquired their B.Sc. degree outside European countries (3 HCRs working in Europe obtained their first degree in the US, however, only one of them was born in the US).

### D. Birthplace of HCRs

Finally, we focus on the data regarding the birthplace of the HCRs (Table 6), which show that the majority of HCRs were born in Europe (37.6%), while 31.5% came from US, and the remaining 27.7% were born in countries in other parts of the world.

*Table 6: Frequencies and percentages of HCRs according to their country of birth*

|  |  | FREQ | (%) |
|---|---|---|---|
| Valid | EU | 129 | 37.6 |
|  | US | 108 | 31.5 |
|  | China/Taiwan | 19 | 5.5 |
|  | Canada | 11 | 3.2 |
|  | Australia | 11 | 3.2 |
|  | Israel | 9 | 2.6 |
|  | India | 9 | 2.6 |
|  | Russia | 8 | 2.3 |
|  | Japan | 5 | 1.5 |
|  | Hong Kong | 4 | 1.2 |
|  | South Africa | 3 | 0.9 |
|  | Argentina | 2 | 0.6 |
|  | New Zealand | 2 | 0.6 |
|  | rest of the world (*) | 12 | 3.5 |
|  | Total | 332 | 96.8 |
| Missing |  | 11 | 3.2 |
| TOTAL |  | 343 | 100.0 |

*(*) 1 HCR for each of Peru, Egypt, Brazil, Mexico, Venezuela, Algeria, Turkey, Chile, Tunisia, Vietnam, Pakistan and Rep of Congo*

In Table 7, a classification of the HCRs with respect to the country of current affiliation and the country of birth is presented. The results are quite similar to the previous results. It is obvious that for the HCRs currently working in the US, less than half were native-born (46.5%), while the vast majority of researchers working in Europe or the rest of the world are native-born citizens (94.7% and 83.3%, respectively). We also see that the movement from Europe to the US (23.9%) heavily outnumbers the opposite movement (1.3%). A more detailed break-down of the percentages is given in Table A7 in the appendix. As observed, the majority of HCRs affiliated with US Institutions, born outside the US and Europe come from China (7.5%), followed by Canada (4%). While the status of a scientist as being highly cited is influenced by his whole career, if we are to accept that these scientists have achieved a potential they had all along, it is clear that the US is doing best in harnessing this potential.

*Table 7: Contingency table between the country of present affiliation and the country of birth of the HCRs*

|  |  |  | Country of Birth of the HCRs | | | TOTAL |
|---|---|---|---|---|---|---|
|  |  |  | US | EU | Rest of the world |  |
| Country of Present Affiliation of the HCRs | US | Count | 105 | 54 | 67 | 226 |
|  |  | % within | 46.5% | 23.9% | 29.6% | 100.0% |
|  | EU | Count | 1 | 72 | 3 | 76 |
|  |  | % within | 1.3% | 94.7% | 3.9% | 100.0% |
|  | Rest of the world | Count | 2 | 3 | 25 | 30 |
|  |  | % within | 6.7% | 10.0% | 83.3% | 100.0% |
| TOTAL |  | Count | 108 | 125 | 95 | 332 |
|  |  | % within | 32.5% | 38.9% | 28.6% | 100.0% |

Generally, the majority of HCRs working in US Universities and Institutions were born elsewhere (121 out of 226 researchers), while exactly the opposite holds true for the rest of the world, where the vast majority of researchers are native-born citizens (see Figure 1).

*Figure 1: Counts of HCRs for the US, European and non-US & European Institutions*

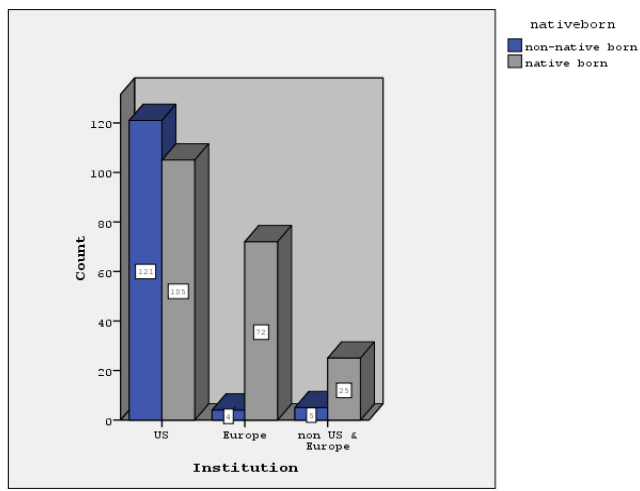

In relation to the movement of HCRs in the early steps of their life, we observe from Table 8 that moving between US, Europe and the rest of the world is retained at the minimum level. Indeed, the vast majority of HCRs complete their B.Sc. studies in their native country (96%, 91.5% and 90%, for US, Europe and the rest of the world, respectively). Still though, the number of HCRs who left Europe (and the rest of the world) in order to study for an undergraduate degree is larger than the number of those who leave the US to go abroad for the same reason.

*Table 8: Contingency table between the country of birth of the HCRs and the country where the first degree of the HCRs was completed*

| | | | Country in which the B.Sc. Degree of the HCRs was obtained | | | TOTAL |
|---|---|---|---|---|---|---|
| | | | US | EU | Rest of the world | |
| Country of Birth of the HCRs | US | Count | 96 | 3 | 1 | 100 |
| | | % within | 96.0% | 3.0% | 1.0% | 100.0% |
| | EU | Count | 7 | 107 | 3 | 117 |
| | | % within | 6.0% | 91.5% | 2.6% | 100.0% |
| | Rest of the world | Count | 6 | 3 | 81 | 90 |
| | | % within | 6.7% | 3.3% | 90.0% | 100.0% |
| TOTAL | | Count | 109 | 113 | 85 | 307 |
| | | % within | 35.5% | 36.8% | 27.7% | 100.0% |

Finally, Table 9 relates the country of undergraduate and Ph.D. studies of the highly cited mathematicians. As we observe, almost all of the researchers who obtained their B.Sc. degree in the US continued their studies there (99.1%). In contrast, a highly significant number of European researchers (20.2%) left Europe to continue their Ph.D. studies in the US, while the majority of the researchers from other countries (59.8%) continued their Ph.D. studies in the US. In total, from the 186 HC researchers that acquired their Ph.D. title in the US, 75 came from European universities and from the rest of the world. A further breakdown can be found in Table A8 of the appendix. By inspection of Table A8, it becomes evident that a significant percentage of the HCRs that completed their Ph.D. studies in the US, had done their undergraduate studies elsewhere, and in particular in Europe (12.4%), China (9.7%), Canada (4.8%), India (3.8%) and Hong Kong (2.2%). It is worth observing that none of the HCRs who did their undergraduate studies in Europe or the US chose to go to another continent for their Ph.D studies.

*Table 9: Contingency table between the country of BS degree and the country of PhD degree of the HCRs*

| | | | Country in which the Ph.D. Degree of the HCRs was obtained | | | TOTAL |
|---|---|---|---|---|---|---|
| | | | US | EU | Rest of the world | |
| Country in which the B.Sc. Degree of the HCRs was obtained | US | Count | 111 | 1 | 0 | 112 |
| | | % within | 99.1% | 0.9% | 0.0% | 100.0% |
| | EU | Count | 23 | 91 | 0 | 114 |
| | | % within | 20.2% | 79.8% | 0.0% | 100.0% |
| | Rest of the world | Count | 52 | 9 | 26 | 87 |
| | | % within | 59.8% | 10.3% | 29.9% | 100.0% |
| TOTAL | | Count | 186 | 101 | 26 | 313 |
| | | % within | 59.4% | 32.3% | 8.3% | 100.0% |

### III. HCRs AND TOP INSTITUTIONS

We now turn to a more detailed investigation, and include the specific university of current affiliation. Table A9 in the appendix lists the Institutions (24 in all) that employ almost half of the HCRs (45.22%) in a total number of 161 Institutions/Universities. It has been reported elsewhere [Bauwens et al. (2007)] that 30.1% of all HCRs in all fields work in the 25 top Institutions. Our findings indicate a much higher concentration of HCRs in top mathematics institutions than in other scientific fields (one might attempt to attribute this to the fact that hiring a top mathematician is less "expensive" for institutions than hiring an experimental scientist). As one may observe, 20 of the top 24 Institutions in Mathematics ranked from the point of view of HCRs are in the USA, while only three are in Europe (University of Oxford, Pierre & Marie Curie University and University of Cambridge) and one is located in Israel (Tel Aviv University).

Observing however, the percentages of native and non-native HCRs in each one of the top Universities it is obvious that for the majority of the US Universities their HCRs come mostly from countries outside the United States. For instance, at Princeton University 8 out of the 10 HCRs come from countries outside the US, while at Rutgers University, all of the HCRs (5) were not born in US (see Figure 2).

*Figure 2: Distribution of native and non-native HCRs across the 24 top ranked Mathematics Departments*

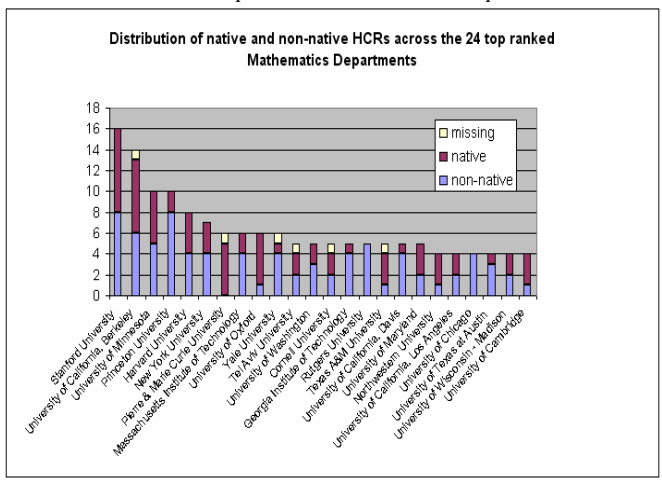

On the other hand, we observe the exact opposite effect when it comes to the three European institutions that complete the table. For example, in Pierre and Marie Curie University and the University of Cambridge, the majority of the HCRs are native-born citizens (5 and 3, respectively), while for the University of Oxford only one out of 5 was born elsewhere. One may argue that the top European Institutions have difficulties in attracting/retaining non-European born HCRs.

We conclude with more general observations regarding the Institutions the HCRs are affiliated with. In Table 11 at the end we present the number of HCRs in mathematics and in all scientific fields in the top ranking Institutions.

The table indicates that the majority of top Institutions as regards their overall performance in number of HCRs working in them, have also high numbers of HCRs in Mathematics. Specifically, 16 out of the 27 top institutions in all disciplines also appear in the top list of the HCRs in Mathematics. Stanford University and the University of California, Berkeley are well ahead of the rest when we look at the number of HCRs in mathematics (4.66% and 4.08% of HCRs in the top ranking Institutions, respectively)[4].

To further investigate the impact of HCRs in mathematics on their Institutions/Universities, we present in Table 10 the proportion of Mathematicians HCRs to the overall number of highly cited researchers in the Institutions. It is evident that the proportion of HCRs in mathematics is higher in Institutions that are mainly (or solely) focused on science, such as the Georgia Institute of Technology or the Pierre & Marie Curie University. It is also of interest to note that in Tel Aviv University there are 5 HCRs in mathematics and 12 HCRs in all Departments.

*Table 10: Percentages of HCRs in Mathematics at the top Institutions*

| Rank | Institution of Affiliation (Mathematics/Statistics) | HCRs in math Departs | HCRs in the University | % of HCRs in math Departs | no of students | Country |
|---|---|---|---|---|---|---|
| 1 | Pierre & Marie Curie University | 6 | 11 | 54.55% | 30.000 | France |
| 2 | Georgia Institute of Technology | 5 | 12 | 41.67% | 18.747 | USA |
| 3 | Tel Aviv University | 5 | 12 | 41.67% | 29.000 | Israel |
| 4 | Texas A&M University | 5 | 22 | 22.73% | 46.540 | USA |
| 5 | New York University | 7 | 31 | 22.58% | 40.870 | USA |
| 6 | University of Minnesota | 10 | 47 | 21.28% | 50.402 | USA |
| 7 | Rutgers University | 5 | 30 | 16.67% | 49.760 | USA |
| 8 | Princeton University | 10 | 68 | 14.71% | 7.334 | USA |
| 9 | University of Oxford | 6 | 45 | 13.33% | 19.486 | UK |
| 10 | University of California, Davis | 5 | 40 | 12.50% | 30.475 | USA |
| 11 | University of Maryland | 5 | 44 | 11.36% | 36.014 | USA |
| 12 | Northwestern University | 4 | 40 | 10.00% | 15.129 | USA |
| 13 | University of Texas at Austin | 4 | 40 | 10.00% | 49.696 | USA |
| 14 | University of California, Berkeley | 14 | 142 | 9.86% | 34.953 | USA |
| 15 | Yale University | 6 | 61 | 9.84% | 16.714 | USA |
| 16 | University of Washington | 5 | 53 | 9.43% | 42.974 | USA |
| 17 | Cornell University | 5 | 54 | 9.26% | 19.800 | USA |
| 18 | Stanford University | 16 | 187 | 8.56% | 14.945 | USA |
| 19 | University of Chicago | 4 | 48 | 8.33% | 14.721 | USA |
| 20 | Massachusetts Institute of Technology | 6 | 76 | 7.89% | 10.220 | USA |
| 21 | University of Cambridge | 4 | 52 | 7.69% | 18.396 | UK |
| 22 | University of Wisconsin - Madison | 4 | 52 | 7.69% | 42.041 | USA |
| 23 | University of California, Los Angeles | 4 | 59 | 6.78% | 36.611 | USA |
| 24 | Harvard University | 8 | 187 | 4.28% | 19.139 | USA |

## IV. Conclusions

The results of the current study verify the widely held belief of a brain drain in mathematics from Europe and the rest of the world to the US, at least among those mathematicians who have become highly cited. Moreover, it provides evidence supporting the view that this brain drain becomes more acute as the career of the HCRs evolves. Focusing within this influential group of mathematicians we see that while only 6% of Europeans moved to the US for their undergraduate studies, the US drained 20% of European bachelors to do a PhD in the US. At the next level, 33.6% of European PhDs were attracted to faculty or research positions in the US.

The situation is worse for the HCRs born outside the US and Europe. The US drained 59.8% of non-European foreign bachelors to do a PhD in the US, while 55.2% of non-European foreign PhDs were attracted to faculty positions in the US.

On the other hand, the retention level of the HCRs in mathematics is high at every level in the US. The US has managed to retain 99% of their bachelors to do their PhDs and 90% of their doctors as faculty members in US Institutions.

These results, combined with other findings in this article, reveal that a significant number of HCRs working in the US has been scientifically "nurtured" elsewhere. The US is able to attract some of the best minds in mathematics from all over the world, and has found the means and conditions to keep them there.

If Europe wants to compete with the US, at least in mathematics, it should follow the example of the US and find ways of not only retaining its best scientists but also of attracting more from other parts of the world, including the US. The European Research Council established recently and the Starting and Advanced Research Grants awarded are certainly a step in the right direction.

---

[4] In cases of ties we have ranked higher the Institution with fewer faculty members. Data on the number of faculty members associated with departments of mathematics/statistics have been collected from each department's web page (data on the number of faculty members of Universities has been collected from wikipedia) (*Wikipedia, The Free Encyclopedia*, http://en.wikipedia.org).

*Table 11: Comparing percentages of HCRs in Mathematics and in all 21 disciplines at the top Institutions*

| Rank | Institution of Affiliation (Mathematics/Statistics) | HCRs | % of HCRs | Country | Institution of Affiliation (All 21 disciplines) | HCRs | % of HCRs | Country | Rank |
|---|---|---|---|---|---|---|---|---|---|
| 1 | Stanford University | 16 | 4.66% | USA | Harvard University | 187 | 3.06% | USA | 1 |
| 2 | University of California, Berkeley | 14 | 4.08% | USA | Stanford University | 142 | 2.33% | USA | 2 |
| 3 | Princeton University | 10 | 2.92% | USA | National Institutes of Health | 136 | 2.23% | USA | 3 |
| 4 | University of Minnesota | 10 | 2.92% | USA | University of California, Berkeley | 87 | 1.43% | USA | 4 |
| 5 | Harvard University | 8 | 2.33% | USA | Massachusetts Institute of Technology | 76 | 1.25% | USA | 6 |
| 6 | New York University | 7 | 2.04% | USA | Max-Planck-Institute | 76 | 1.25% | Germany | 5 |
| 7 | University of Oxford | 6 | 1.75% | UK | Princeton University | 68 | 1.11% | USA | 8 |
| 8 | Yale University | 6 | 1.75% | USA | University of Michigan | 68 | 1.11% | USA | 7 |
| 9 | Massachusetts Institute of Technology | 6 | 1.75% | USA | University of California, San Diego | 66 | 1.08% | USA | 9 |
| 10 | Pierre & Marie Curie University | 6 | 1.75% | France | University of Pennsylvania | 64 | 1.05% | USA | 10 |
| 11 | Cornell University | 5 | 1.46% | USA | California Institute of Technology | 61 | 1.00% | USA | 12 |
| 12 | University of California, Davis | 5 | 1.46% | USA | Yale University | 61 | 1.00% | USA | 11 |
| 13 | University of Maryland | 5 | 1.46% | USA | University of California, Los Angeles | 59 | 0.97% | USA | 13 |
| 14 | University of Washington | 5 | 1.46% | USA | University of California, San Francisco | 54 | 0.88% | USA | 14 |
| 15 | Georgia Institute of Technology | 5 | 1.46% | USA | Cornell University | 54 | 0.88% | USA | 15 |
| 16 | Rutgers University | 5 | 1.46% | USA | University of Washington | 53 | 0.87% | USA | 16 |
| 17 | Tel Aviv University | 5 | 1.46% | Israel | University of Wisconsin - Madison | 52 | 0.85% | USA | 17 |
| 18 | Texas A&M University | 5 | 1.46% | USA | Columbia University | 52 | 0.85% | USA | 18 |
| 19 | University of Cambridge | 4 | 1.17% | UK | University of Cambridge | 51 | 0.84% | UK | 19 |
| 20 | University of Chicago | 4 | 1.17% | USA | University of Chicago | 48 | 0.79% | USA | 20 |
| 21 | Northwestern University | 4 | 1.17% | USA | University of Minnesota | 47 | 0.77% | USA | 21 |
| 22 | University of Wisconsin - Madison | 4 | 1.17% | USA | University of Oxford | 45 | 0.74% | UK | 22 |
| 23 | University of California, Los Angeles | 4 | 1.17% | USA | University of Maryland | 44 | 0.72% | USA | 23 |
| 24 | University of Texas at Austin | 4 | 1.17% | USA | NASA | 43 | 0.70% | USA | 24 |
|  |  |  |  |  | Duke University | 41 | 0.67% | USA | 25 |
|  |  |  |  |  | University of California, Davis | 40 | 0.66% | USA | 26 |
|  |  |  |  |  | Northwestern University | 40 | 0.66% | USA | 27 |

**APPENDIX**

*Table A1: Numbers of HCRs in all 21 disciplines according to their present affiliation.*

| Country of present affiliation | Number of HCRs | Percentage of HCRs |
|---|---|---|
| United States | 4007 | 65.66% |
| United Kingdom | 464 | 7.60% |
| Germany | 262 | 4.29% |
| Japan | 256 | 4.19% |
| Canada | 185 | 3.03% |
| France | 163 | 2.67% |
| Switzerland | 113 | 1.85% |
| Australia | 109 | 1.79% |
| Netherlands | 97 | 1.59% |
| Italy | 81 | 1.33% |
| Sweden | 62 | 1.02% |
| Israel | 48 | 0.79% |
| Belgium | 39 | 0.64% |
| Denmark | 31 | 0.51% |
| Spain | 22 | 0.36% |
| Peoples Rep China | 20 | 0.33% |
| New Zealand | 18 | 0.29% |
| Finland | 17 | 0.28% |
| Austria | 13 | 0.21% |
| Norway | 13 | 0.21% |
| India | 11 | 0.18% |
| Taiwan | 9 | 0.15% |
| Ireland | 8 | 0.13% |
| South Africa | 7 | 0.11% |
| Hungary | 6 | 0.10% |
| Russia | 6 | 0.10% |
| Brazil | 5 | 0.08% |
| Greece | 5 | 0.08% |
| Chile | 4 | 0.07% |
| Singapore | 4 | 0.07% |
| Mexico | 3 | 0.05% |
| Republic of Korea | 3 | 0.05% |
| Panama | 2 | 0.03% |
| Poland | 2 | 0.03% |
| Algeria | 1 | 0.02% |
| Hong Kong | 1 | 0.02% |
| Iran | 1 | 0.02% |
| Pakistan | 1 | 0.02% |
| Philippines | 1 | 0.02% |
| Portugal | 1 | 0.02% |
| Romania | 1 | 0.02% |
| Turkey | 1 | 0,02% |
| **TOTAL** | **6103** | **100%** |

*Table A2: Numbers of HCRs in all 21 disciplines according to their present affiliation.*

| Country of present affiliation | Number of HCRs | Percentage of HCRs |
|---|---|---|
| United States | 4007 | 65.66% |
| EU | 1400 | 22.94% |
| Rest of the world | 696 | 11.40% |
| **TOTAL** | **6103** | **100%** |

*Table A3: Numbers of HCRs in the field of Mathematics according to their present affiliation.*

| Country of present affiliation | Number of HCRs | Percentage of HCRs |
|---|---|---|
| United States | 234 | 68.22% |
| United Kingdom | 24 | 7.00% |
| France | 22 | 6.41% |
| Germany | 9 | 2.62% |
| Israel | 8 | 2.33% |
| Australia | 6 | 1.75% |
| Canada | 6 | 1.75% |
| Japan | 5 | 1.46% |
| Denmark | 4 | 1.17% |
| Italy | 4 | 1.17% |
| Netherlands | 4 | 1.17% |
| Spain | 4 | 1.17% |
| Switzerland | 3 | 0.87% |
| Hungary | 2 | 0.58% |
| Peoples Rep of China | 2 | 0.58% |
| Belgium | 1 | 0.29% |
| India | 1 | 0.29% |
| Singapore | 1 | 0.29% |
| Sweden | 1 | 0.29% |
| Taiwan | 1 | 0.29% |
| Turkey | 1 | 0.29% |
| **TOTAL** | **343** | **100,00%** |

*Table A4: Distribution of HCRs in all 21 disciplines according to their present affiliation and discipline.*

| Discipline | Country of present affiliation | | | TOTAL |
|---|---|---|---|---|
| | US | EU | Rest of the world | |
| Agricultural Sciences | 118 | 88 | 73 | 279 |
| | 42.3% | 31.5% | 26.2% | 100.0% |
| Biology and Biochemistry | 141 | 43 | 41 | 225 |
| | 62.7% | 19.1% | 18.2% | 100.0% |
| Chemistry | 143 | 72 | 35 | 250 |
| | 57.2% | 28.8% | 14.0% | 100.0% |
| Clinical Medicine | 166 | 41 | 12 | 219 |
| | 75.8% | 18.7% | 5.5% | 100.0% |
| Computer Science | 241 | 46 | 39 | 326 |
| | 73.9% | 14.1% | 12.0% | 100.0% |
| Ecology-Environment | 201 | 75 | 36 | 312 |
| | 64.4% | 24.0% | 11.5% | 100.0% |
| Economics-Business | 268 | 26 | 17 | 311 |
| | 86.2% | 8.4% | 5.5% | 100.0% |
| Engineering | 142 | 39 | 30 | 211 |
| | 67.3% | 18.5% | 14.2% | 100.0% |
| Geosciences | 219 | 73 | 24 | 316 |
| | 69.3% | 23.1% | 7.6% | 100.0% |
| Immunology | 209 | 84 | 35 | 328 |
| | 63.7% | 25.6% | 10.7% | 100.0% |
| Materials Science | 163 | 55 | 55 | 273 |
| | 59.7% | 20.1% | 20.1% | 100.0% |
| Mathematics | 225 | 78 | 31 | 334 |
| | 67.4% | 23.4% | 9.3% | 100.0% |
| Microbiology | 215 | 96 | 24 | 335 |
| | 64.2% | 28.7% | 7.2% | 100.0% |
| Molecular Biology and Genetics | 215 | 65 | 21 | 301 |
| | 71.4% | 21.6% | 7.0% | 100.0% |
| Neuroscience | 190 | 85 | 22 | 297 |
| | 64.0% | 28.6% | 7.4% | 100.0% |
| Pharmacology | 94 | 123 | 46 | 263 |
| | 35.7% | 46.8% | 17.5% | 100.0% |
| Physics | 160 | 91 | 37 | 288 |
| | 55.6% | 31.6% | 12.8% | 100.0% |
| Plant and Animal Science | 148 | 101 | 56 | 305 |
| | 48.5% | 33.1% | 18.4% | 100.0% |
| Psychology-Psychiatry | 229 | 24 | 13 | 266 |
| | 86.1% | 9.0% | 4.9% | 100.0% |
| Social Sciences, General | 296 | 12 | 10 | 318 |
| | 93.1% | 3.8% | 3.1% | 100.0% |
| Space Sciences | 224 | 83 | 39 | 346 |
| | 64.7% | 24.0% | 11.3% | 100.0% |
| **TOTAL** | **4.007** | **1.400** | **696** | **6.103** |
| | **65.7%** | **22.9%** | **11.4%** | **100.0%** |

***Table A5:*** *Contingency table between the country of present affiliation and the country where the first degree was completed in the field of mathematics*

| | | Country in which the B.Sc. Degree was obtained | | | | | | | | | | |
|---|---|---|---|---|---|---|---|---|---|---|---|---|
| | | US | EU | India | Canada | Russia | Israel | China-Taiwan | Australia | Japan | Turkey | Argentina |
| Country of Present Affiliation | US | 107 | 50 | 7 | 11 | 4 | 2 | 16 | 5 | 0 | 0 | 1 |
| | | 49.1% | 22.9% | 3.2% | 5.0% | 1.8% | 0.9% | 7.3% | 2.3% | 0.0% | 0.0% | 0.5% |
| | EU | 3 | 62 | 0 | 0 | 2 | 0 | 0 | 0 | 0 | 0 | 0 |
| | | 4.5% | 92.5% | 0.0% | 0.0% | 3.0% | 0.0% | 0.0% | 0.0% | 0.0% | 0.0% | 0.0% |
| | India | 0 | 0 | 1 | 0 | 0 | 0 | 0 | 0 | 0 | 0 | 0 |
| | | 0.0% | 0.0% | 100.0% | 0.0% | 0.0% | 0.0% | 0.0% | 0.0% | 0.0% | 0.0% | 0.0% |
| | Canada | 1 | 0 | 1 | 3 | 0 | 0 | 0 | 1 | 0 | 0 | 0 |
| | | 16.7% | 0.0% | 16.7% | 50.0% | 0.0% | 0.0% | 0.0% | 16.7% | 0.0% | 0.0% | 0.0% |
| | Israel | 1 | 0 | 0 | 0 | 1 | 4 | 0 | 0 | 0 | 0 | 0 |
| | | 16.7% | 0.0% | 0.0% | 0.0% | 16.7% | 66.7% | 0.0% | 0.0% | 0.0% | 0.0% | 0.0% |
| | China-Taiwan | 0 | 0 | 0 | 0 | 0 | 0 | 2 | 0 | 0 | 0 | 0 |
| | | 0.0% | 0.0% | 0.0% | 0.0% | 0.0% | 0.0% | 66.7% | 0.0% | 0.0% | 0.0% | 0.0% |
| | Australia | 0 | 1 | 0 | 0 | 0 | 0 | 0 | 5 | 0 | 0 | 0 |
| | | 0.0% | 16.7% | 0.0% | 0.0% | 0.0% | 0.0% | 0.0% | 83.3% | 0.0% | 0.0% | 0.0% |
| | Japan | 0 | 0 | 0 | 0 | 0 | 0 | 0 | 0 | 4 | 0 | 0 |
| | | 0.0% | 0.0% | 0.0% | 0.0% | 0.0% | 0.0% | 0.0% | 0.0% | 100.0% | 0.0% | 0.0% |
| | Singapore | 0 | 1 | 0 | 0 | 0 | 0 | 0 | 0 | 0 | 0 | 0 |
| | | 0.0% | 100.0% | 0.0% | 0.0% | 0.0% | 0.0% | 0.0% | 0.0% | 0.0% | 0.0% | 0.0% |
| | Turkey | 0 | 0 | 0 | 0 | 0 | 0 | 0 | 0 | 0 | 1 | 0 |
| | | 0.0% | 0.0% | 0.0% | 0.0% | 0.0% | 0.0% | 0.0% | 0.0% | 0.0% | 100.0% | 0.0% |
| | TOTAL | 112 | 114 | 9 | 14 | 7 | 6 | 18 | 11 | 4 | 1 | 1 |
| | | 35.8% | 36.4% | 2.9% | 4.5% | 2.2% | 1.9% | 5.8% | 3.5% | 1.3% | 0.3% | 0.3% |

| | | Country in which the B.Sc. Degree was obtained | | | | | | | | | TOTAL |
|---|---|---|---|---|---|---|---|---|---|---|---|
| | | Hong Kong | Peru | South Africa | Egypt | Brazil | Mexico | New Zealand | Venezuela | Algeria | Chile | |
| Country of Present Affiliation | US | 3 | 1 | 3 | 1 | 1 | 1 | 2 | 1 | 1 | 1 | 218 |
| | | 1.4% | 0.5% | 1.4% | 0.5% | 0.5% | 0.5% | 0.9% | 0.5% | 0.5% | 0.5% | 100.0% |
| | EU | 0 | 0 | 0 | 0 | 0 | 0 | 0 | 0 | 0 | 0 | 67 |
| | | 0.0% | 0.0% | 0.0% | 0.0% | 0.0% | 0.0% | 0.0% | 0.0% | 0.0% | 0.0% | 100.0% |
| | India | 0 | 0 | 0 | 0 | 0 | 0 | 0 | 0 | 0 | 0 | 1 |
| | | 0.0% | 0.0% | 0.0% | 0.0% | 0.0% | 0.0% | 0.0% | 0.0% | 0.0% | 0.0% | 100.0% |
| | Canada | 0 | 0 | 0 | 0 | 0 | 0 | 0 | 0 | 0 | 0 | 6 |
| | | 0.0% | 0.0% | 0.0% | 0.0% | 0.0% | 0.0% | 0.0% | 0.0% | 0.0% | 0.0% | 100.0% |
| | Israel | 0 | 0 | 0 | 0 | 0 | 0 | 0 | 0 | 0 | 0 | 6 |
| | | 0.0% | 0.0% | 0.0% | 0.0% | 0.0% | 0.0% | 0.0% | 0.0% | 0.0% | 0.0% | 100.0% |
| | China-Taiwan | 1 | 0 | 0 | 0 | 0 | 0 | 0 | 0 | 0 | 0 | 3 |
| | | 33.3% | 0.0% | 0.0% | 0.0% | 0.0% | 0.0% | 0.0% | 0.0% | 0.0% | 0.0% | 100.0% |
| | Australia | 0 | 0 | 0 | 0 | 0 | 0 | 0 | 0 | 0 | 0 | 6 |
| | | 0.0% | 0.0% | 0.0% | 0.0% | 0.0% | 0.0% | 0.0% | 0.0% | 0.0% | 0.0% | 100.0% |
| | Japan | 0 | 0 | 0 | 0 | 0 | 0 | 0 | 0 | 0 | 0 | 4 |
| | | 0.0% | 0.0% | 0.0% | 0.0% | 0.0% | 0.0% | 0.0% | 0.0% | 0.0% | 0.0% | 100.0% |
| | Singapore | 0 | 0 | 0 | 0 | 0 | 0 | 0 | 0 | 0 | 0 | 1 |
| | | 0.0% | 0.0% | 0.0% | 0.0% | 0.0% | 0.0% | 0.0% | 0.0% | 0.0% | 0.0% | 100.0% |
| | Turkey | 0 | 0 | 0 | 0 | 0 | 0 | 0 | 0 | 0 | 0 | 1 |
| | | 0.0% | 0.0% | 0.0% | 0.0% | 0.0% | 0.0% | 0.0% | 0.0% | 0.0% | 0.0% | 100.0% |
| | TOTAL | 4 | 1 | 3 | 1 | 1 | 1 | 2 | 1 | 1 | 1 | 313 |
| | | 1.3% | 0.3% | 1.0% | 0.3% | 0.3% | 0.3% | 0.6% | 0.3% | 0.3% | 0.3% | 100.0% |

*Table A6:* Contingency table between the country of present affiliation and the country of Ph.D. degree in the field of mathematics

| | | Country of Present Affiliation | | | | | | | | | | |
|---|---|---|---|---|---|---|---|---|---|---|---|---|
| | | US | EU | India | Canada | Israel | China-Taiwan | Australia | Japan | Singapore | Turkey | TOTAL |
| Country in which the Ph.D. Degree was obtained | US | 180 | 6 | 1 | 3 | 3 | 3 | 1 | 0 | 0 | 1 | 198 |
| | | 90.9% | 3.0% | 0.5% | 1.5% | 1.5% | 1.5% | 0.5% | 0.0% | 0.0% | 0.5% | 100.0% |
| | EU | 37 | 65 | 0 | 3 | 0 | 0 | 4 | 0 | 1 | 0 | 110 |
| | | 33.6% | 59.1% | 0.0% | 2.7% | 0.0% | 0.0% | 3.6% | 0.0% | 0.9% | 0.0% | 100.0% |
| | India | 2 | 0 | 0 | 0 | 0 | 0 | 0 | 0 | 0 | 0 | 2 |
| | | 100.0% | 0.0% | 0.0% | 0.0% | 0.0% | 0.0% | 0.0% | 0.0% | 0.0% | 0.0% | 100.0% |
| | Canada | 6 | 0 | 0 | 0 | 0 | 0 | 0 | 0 | 0 | 0 | 6 |
| | | 100.0% | 0.0% | 0.0% | 0.0% | 0.0% | 0.0% | 0.0% | 0.0% | 0.0% | 0.0% | 100.0% |
| | Russia | 2 | 2 | 0 | 0 | 1 | 0 | 0 | 0 | 0 | 0 | 5 |
| | | 40.0% | 40.0% | 0.0% | 0.0% | 20.0% | 0.0% | 0.0% | 0.0% | 0.0% | 0.0% | 100.0% |
| | Israel | 3 | 0 | 0 | 0 | 4 | 0 | 0 | 0 | 0 | 0 | 7 |
| | | 42.9% | 0.0% | 0.0% | 0.0% | 57.1% | 0.0% | 0.0% | 0.0% | 0.0% | 0.0% | 100.0% |
| | Australia | 1 | 0 | 0 | 0 | 0 | 0 | 1 | 0 | 0 | 0 | 2 |
| | | 50.0% | 0.0% | 0.0% | 0.0% | 0.0% | 0.0% | 50.0% | 0.0% | 0.0% | 0.0% | 100.0% |
| | Japan | 0 | 0 | 0 | 0 | 0 | 0 | 0 | 5 | 0 | 0 | 5 |
| | | 0.0% | 0.0% | 0.0% | 0.0% | 0.0% | 0.0% | 0.0% | 100.0% | 0.0% | 0.0% | 100.0% |
| | Argentina | 1 | 0 | 0 | 0 | 0 | 0 | 0 | 0 | 0 | 0 | 1 |
| | | 100.0% | 0.0% | 0.0% | 0.0% | 0.0% | 0.0% | 0.0% | 0.0% | 0.0% | 0.0% | 100.0% |
| | South Africa | 1 | 0 | 0 | 0 | 0 | 0 | 0 | 0 | 0 | 0 | 1 |
| | | 100.0% | 0.0% | 0.0% | 0.0% | 0.0% | 0.0% | 0.0% | 0.0% | 0.0% | 0.0% | 100.0% |
| TOTAL | | 233 | 73 | 1 | 6 | 8 | 3 | 6 | 5 | 1 | 1 | 337 |
| | | 69.1% | 21.7% | 0.3% | 1.8% | 2.4% | 0.9% | 1.8% | 1.5% | 0.3% | 0.3% | 100.0% |

*Table A7: Contingency table between the country of present affiliation and the country of birth in the field of mathematics*

| | | Country of Birth | | | | | | | | | | | | |
|---|---|---|---|---|---|---|---|---|---|---|---|---|---|---|
| | | US | EU | India | Canada | Russia | Israel | China-Taiwan | Australia | Japan | Turkey | Argentina | Hong Kong | Peru |
| Country of Present Affiliation | US | 105 | 54 | 7 | 9 | 5 | 4 | 17 | 5 | 0 | 0 | 2 | 3 | 1 |
| | | 46.5% | 23.9% | 3.1% | 4.0% | 2.2% | 1.8% | 7.5% | 2.2% | 0.0% | 0.0% | 0.9% | 1.3% | 0.4% |
| | EU | 1 | 72 | 0 | 0 | 2 | 0 | 0 | 0 | 0 | 0 | 0 | 0 | 0 |
| | | 1.3% | 94.7% | 0.0% | 0.0% | 2.6% | 0.0% | 0.0% | 0.0% | 0.0% | 0.0% | 0.0% | 0.0% | 0.0% |
| | India | 0 | 0 | 1 | 0 | 0 | 0 | 0 | 0 | 0 | 0 | 0 | 0 | 0 |
| | | 0.0% | 0.0% | 100.0% | 0.0% | 0.0% | 0.0% | 0.0% | 0.0% | 0.0% | 0.0% | 0.0% | 0.0% | 0.0% |
| | Canada | 1 | 1 | 1 | 2 | 0 | 0 | 0 | 1 | 0 | 0 | 0 | 0 | 0 |
| | | 16.7% | 16.7% | 16.7% | 33.3% | 0.0% | 0.0% | 0.0% | 16.7% | 0.0% | 0.0% | 0.0% | 0.0% | 0.0% |
| | Israel | 1 | 0 | 0 | 0 | 1 | 5 | 0 | 0 | 0 | 0 | 0 | 0 | 0 |
| | | 14.3% | 0.0% | 0.0% | 0.0% | 14.3% | 71.4% | 0.0% | 0.0% | 0.0% | 0.0% | 0.0% | 0.0% | 0.0% |
| | China-Taiwan | 0 | 0 | 0 | 0 | 0 | 0 | 2 | 0 | 0 | 0 | 0 | 1 | 0 |
| | | 0.0% | 0.0% | 0.0% | 0.0% | 0.0% | 0.0% | 66.7% | 0.0% | 0.0% | 0.0% | 0.0% | 33.3% | 0.0% |
| | Australia | 0 | 1 | 0 | 0 | 0 | 0 | 0 | 5 | 0 | 0 | 0 | 0 | 0 |
| | | 0.0% | 16.7% | 0.0% | 0.0% | 0.0% | 0.0% | 0.0% | 83.3% | 0.0% | 0.0% | 0.0% | 0.0% | 0.0% |
| | Japan | 0 | 0 | 0 | 0 | 0 | 0 | 0 | 0 | 5 | 0 | 0 | 0 | 0 |
| | | 0.0% | 0.0% | 0.0% | 0.0% | 0.0% | 0.0% | 0.0% | 0.0% | 100.0% | 0.0% | 0.0% | 0.0% | 0.0% |
| | Singapore | 0 | 1 | 0 | 0 | 0 | 0 | 0 | 0 | 0 | 0 | 0 | 0 | 0 |
| | | 0.0% | 100.0% | 0.0% | 0.0% | 0.0% | 0.0% | 0.0% | 0.0% | 0.0% | 0.0% | 0.0% | 0.0% | 0.0% |
| | Turkey | 0 | 0 | 0 | 0 | 0 | 0 | 0 | 0 | 0 | 1 | 0 | 0 | 0 |
| | | 0.0% | 0.0% | 0.0% | 0.0% | 0.0% | 0.0% | 0.0% | 0.0% | 0.0% | 100.0% | 0.0% | 0.0% | 0.0% |
| | TOTAL | 108 | 129 | 9 | 11 | 8 | 9 | 19 | 11 | 5 | 1 | 2 | 4 | 1 |
| | | 32.5% | 38.9% | 2.7% | 3.3% | 2.4% | 2.7% | 5.7% | 3.3% | 1.5% | 0.3% | 0.6% | 1.2% | 0.3% |

| | | Country of Birth | | | | | | | | | | | | |
|---|---|---|---|---|---|---|---|---|---|---|---|---|---|---|
| | | South Africa | Egypt | Brazil | Mexico | New Zealand | Venezuela | Algeria | Chile | Tunisia | Vietnam | Pakistan | Rep of Congo | TOTAL |
| Country of Present Affiliation | US | 3 | 1 | 1 | 1 | 2 | 1 | 1 | 1 | 1 | 1 | 1 | 0 | 226 |
| | | 1.3% | 0.4% | 0.4% | 0.4% | 0.9% | 0.4% | 0.4% | 0.4% | 0.4% | 0.4% | 0.4% | 0.0% | 100.0% |
| | EU | 0 | 0 | 0 | 0 | 0 | 0 | 0 | 0 | 0 | 0 | 0 | 1 | 76 |
| | | 0.0% | 0.0% | 0.0% | 0.0% | 0.0% | 0.0% | 0.0% | 0.0% | 0.0% | 0.0% | 0.0% | 1.3% | 100.0% |
| | India | 0 | 0 | 0 | 0 | 0 | 0 | 0 | 0 | 0 | 0 | 0 | 0 | 1 |
| | | 0.0% | 0.0% | 0.0% | 0.0% | 0.0% | 0.0% | 0.0% | 0.0% | 0.0% | 0.0% | 0.0% | 0.0% | 100.0% |
| | Canada | 0 | 0 | 0 | 0 | 0 | 0 | 0 | 0 | 0 | 0 | 0 | 0 | 6 |
| | | 0.0% | 0.0% | 0.0% | 0.0% | 0.0% | 0.0% | 0.0% | 0.0% | 0.0% | 0.0% | 0.0% | 0.0% | 100.0% |
| | Israel | 0 | 0 | 0 | 0 | 0 | 0 | 0 | 0 | 0 | 0 | 0 | 0 | 7 |
| | | 0.0% | 0.0% | 0.0% | 0.0% | 0.0% | 0.0% | 0.0% | 0.0% | 0.0% | 0.0% | 0.0% | 0.0% | 100.0% |
| | China-Taiwan | 0 | 0 | 0 | 0 | 0 | 0 | 0 | 0 | 0 | 0 | 0 | 0 | 3 |
| | | 0.0% | 0.0% | 0.0% | 0.0% | 0.0% | 0.0% | 0.0% | 0.0% | 0.0% | 0.0% | 0.0% | 0.0% | 100.0% |
| | Australia | 0 | 0 | 0 | 0 | 0 | 0 | 0 | 0 | 0 | 0 | 0 | 0 | 6 |
| | | 0.0% | 0.0% | 0.0% | 0.0% | 0.0% | 0.0% | 0.0% | 0.0% | 0.0% | 0.0% | 0.0% | 0.0% | 100.0% |
| | Japan | 0 | 0 | 0 | 0 | 0 | 0 | 0 | 0 | 0 | 0 | 0 | 0 | 5 |
| | | 0.0% | 0.0% | 0.0% | 0.0% | 0.0% | 0.0% | 0.0% | 0.0% | 0.0% | 0.0% | 0.0% | 0.0% | 100.0% |
| | Singapore | 0 | 0 | 0 | 0 | 0 | 0 | 0 | 0 | 0 | 0 | 0 | 0 | 1 |
| | | 0.0% | 0.0% | 0.0% | 0.0% | 0.0% | 0.0% | 0.0% | 0.0% | 0.0% | 0.0% | 0.0% | 0.0% | 100.0% |
| | Turkey | 0 | 0 | 0 | 0 | 0 | 0 | 0 | 0 | 0 | 0 | 0 | 0 | 1 |
| | | 0.0% | 0.0% | 0.0% | 0.0% | 0.0% | 0.0% | 0.0% | 0.0% | 0.0% | 0.0% | 0.0% | 0.0% | 100.0% |
| | TOTAL | 3 | 1 | 1 | 1 | 2 | 1 | 1 | 1 | 1 | 1 | 1 | 1 | 332 |
| | | 0.9% | 0.3% | 0.3% | 0.3% | 0.6% | 0.3% | 0.3% | 0.3% | 0.3% | 0.3% | 0.3% | 0.3% | 100.0% |

*Table A8: Contingency table between the country of BS degree and the country of PhD degree in the field of mathematics*

| | | Country in which the Ph.D. Degree was obtained | | | | | | | | | | |
|---|---|---|---|---|---|---|---|---|---|---|---|---|
| | | US | EU | India | Canada | Russia | Israel | Australia | Japan | Argentina | South Africa | TOTAL |
| Country in which the B.Sc. Degree was obtained | US | 111 | 1 | 0 | 0 | 0 | 0 | 0 | 0 | 0 | 0 | 112 |
| | | 59.7% | 1.0% | 0.0% | 0.0% | 0.0% | 0.0% | 0.0% | 0.0% | 0.0% | 0.0% | 35.8% |
| | EU | 23 | 91 | 0 | 0 | 0 | 0 | 0 | 0 | 0 | 0 | 114 |
| | | 12.4% | 90.1% | 0.0% | 0.0% | 0.0% | 0.0% | 0.0% | 0.0% | 0.0% | 0.0% | 36.4% |
| | India | 7 | 0 | 2 | 0 | 0 | 0 | 0 | 0 | 0 | 0 | 9 |
| | | 3.8% | 0.0% | 100.0% | 0.0% | 0.0% | 0.0% | 0.0% | 0.0% | 0.0% | 0.0% | 2.9% |
| | Canada | 9 | 2 | 0 | 3 | 0 | 0 | 0 | 0 | 0 | 0 | 14 |
| | | 4.8% | 2.0% | 0.0% | 60.0% | 0.0% | 0.0% | 0.0% | 0.0% | 0.0% | 0.0% | 4.5% |
| | Russia | 2 | 0 | 0 | 0 | 5 | 0 | 0 | 0 | 0 | 0 | 7 |
| | | 1.1% | 0.0% | 0.0% | 0.0% | 100.0% | 0.0% | 0.0% | 0.0% | 0.0% | 0.0% | 2.2% |
| | Israel | 0 | 0 | 0 | 0 | 0 | 6 | 0 | 0 | 0 | 0 | 6 |
| | | 0.0% | 0.0% | 0.0% | 0.0% | 0.0% | 100.0% | 0.0% | 0.0% | 0.0% | 0.0% | 1.9% |
| | China-Taiwan | 18 | 0 | 0 | 0 | 0 | 0 | 0 | 0 | 0 | 0 | 18 |
| | | 9.7% | 0.0% | 0.0% | 0.0% | 0.0% | 0.0% | 0.0% | 0.0% | 0.0% | 0.0% | 5.8% |
| | Australia | 3 | 5 | 0 | 1 | 0 | 0 | 2 | 0 | 0 | 0 | 11 |
| | | 1.6% | 5.0% | 0.0% | 20.0% | 0.0% | 0.0% | 100.0% | 0.0% | 0.0% | 0.0% | 3.5% |
| | Japan | 0 | 0 | 0 | 0 | 0 | 0 | 0 | 4 | 0 | 0 | 4 |
| | | 0.0% | 0.0% | 0.0% | 0.0% | 0.0% | 0.0% | 0.0% | 100.0% | 0.0% | 0.0% | 1.3% |
| | Turkey | 1 | 0 | 0 | 0 | 0 | 0 | 0 | 0 | 0 | 0 | 1 |
| | | 0.5% | 0.0% | 0.0% | 0.0% | 0.0% | 0.0% | 0.0% | 0.0% | 0.0% | 0.0% | 0.3% |
| | Argentina | 0 | 0 | 0 | 0 | 0 | 0 | 0 | 0 | 1 | 0 | 1 |
| | | 0.0% | 0.0% | 0.0% | 0.0% | 0.0% | 0.0% | 0.0% | 0.0% | 100.0% | 0.0% | 0.3% |
| | Hong Kong | 4 | 0 | 0 | 0 | 0 | 0 | 0 | 0 | 0 | 0 | 4 |
| | | 2.2% | 0.0% | 0.0% | 0.0% | 0.0% | 0.0% | 0.0% | 0.0% | 0.0% | 0.0% | 1.3% |
| | Peru | 1 | 0 | 0 | 0 | 0 | 0 | 0 | 0 | 0 | 0 | 1 |
| | | 0.5% | 0.0% | 0.0% | 0.0% | 0.0% | 0.0% | 0.0% | 0.0% | 0.0% | 0.0% | 0.3% |
| | South Africa | 2 | 0 | 0 | 0 | 0 | 0 | 0 | 0 | 0 | 1 | 3 |
| | | 1.1% | 0.0% | 0.0% | 0.0% | 0.0% | 0.0% | 0.0% | 0.0% | 0.0% | 100.0% | 1.0% |
| | Egypt | 0 | 0 | 0 | 1 | 0 | 0 | 0 | 0 | 0 | 0 | 1 |
| | | 0.0% | 0.0% | 0.0% | 20.0% | 0.0% | 0.0% | 0.0% | 0.0% | 0.0% | 0.0% | 0.3% |
| | Brazil | 1 | 0 | 0 | 0 | 0 | 0 | 0 | 0 | 0 | 0 | 1 |
| | | 0.5% | 0.0% | 0.0% | 0.0% | 0.0% | 0.0% | 0.0% | 0.0% | 0.0% | 0.0% | 0.3% |
| | Mexico | 1 | 0 | 0 | 0 | 0 | 0 | 0 | 0 | 0 | 0 | 1 |
| | | 0.5% | 0.0% | 0.0% | 0.0% | 0.0% | 0.0% | 0.0% | 0.0% | 0.0% | 0.0% | 0.3% |
| | New Zealand | 1 | 1 | 0 | 0 | 0 | 0 | 0 | 0 | 0 | 0 | 2 |
| | | 0.5% | 1.0% | 0.0% | 0.0% | 0.0% | 0.0% | 0.0% | 0.0% | 0.0% | 0.0% | 0.6% |
| | Venezuela | 1 | 0 | 0 | 0 | 0 | 0 | 0 | 0 | 0 | 0 | 1 |
| | | 0.5% | 0.0% | 0.0% | 0.0% | 0.0% | 0.0% | 0.0% | 0.0% | 0.0% | 0.0% | 0.3% |
| | Algeria | 0 | 1 | 0 | 0 | 0 | 0 | 0 | 0 | 0 | 0 | 1 |
| | | 0.0% | 1.0% | 0.0% | 0.0% | 0.0% | 0.0% | 0.0% | 0.0% | 0.0% | 0.0% | 0.3% |
| | Chile | 1 | 0 | 0 | 0 | 0 | 0 | 0 | 0 | 0 | 0 | 1 |
| | | 0.5% | 0.0% | 0.0% | 0.0% | 0.0% | 0.0% | 0.0% | 0.0% | 0.0% | 0.0% | 0.3% |
| TOTAL | | 186 | 101 | 2 | 5 | 5 | 6 | 2 | 4 | 1 | 1 | 313 |
| | | 100.0% | 100.0% | 100.0% | 100.0% | 100.0% | 100.0% | 100.0% | 100.0% | 100.0% | 100.0% | 100.0% |

*Table A9: Top Institutions in the field of Mathematics with reference to HCRs*

| Institution of Affiliation | HCRs | % of HCRs | non-native HCRs | % of non-native HCRs | native HCRs | % of native HCRs | BSs acquired in same country | % of BSs acquired in same country | BSs acquired elsewere | % of BSs acquired elsewere | PhDs acquired in same country | % of PhDs acquired in same country | PhDs acquired elsewere | % of PhDs acquired elsewere | Country |
|---|---|---|---|---|---|---|---|---|---|---|---|---|---|---|---|
| Stanford University | 16 | 4.66% | 8 | 50.0% | 8 | 50.0% | 8 | 50.0% | 8 | 50.0% | 16 | 100.0% | 0 | 0.0% | USA |
| University of California. Berkeley (*) | 14 | 4.08% | 6 | 42.9% | 7 | 50.0% | 7 | 50.0% | 5 | 35.7% | 11 | 78.6% | 3 | 21.4% | USA |
| University of Minnesota | 10 | 2.92% | 5 | 50.0% | 5 | 50.0% | 6 | 60.0% | 3 | 30.0% | 8 | 80.0% | 2 | 20.0% | USA |
| Princeton University | 10 | 2.92% | 8 | 80.0% | 2 | 20.0% | 3 | 30.0% | 7 | 70.0% | 5 | 50.0% | 5 | 50.0% | USA |
| Harvard University | 8 | 2.33% | 4 | 50.0% | 4 | 50.0% | 4 | 50.0% | 4 | 50.0% | 8 | 100.0% | 0 | 0.0% | USA |
| New York University | 7 | 2.04% | 4 | 57.1% | 3 | 42.9% | 4 | 57.1% | 3 | 42.9% | 6 | 85.7% | 1 | 14.3% | USA |
| Pierre & Marie Curie University (*) | 6 | 1.75% | 0 | 0.0% | 5 | 83.3% | 4 | 66.7% | 0 | 0.0% | 3 | 50.0% | 2 | 33.3% | France |
| Massachusetts Institute of Technology | 6 | 1.75% | 4 | 66.7% | 2 | 33.3% | 1 | 16.7% | 5 | 83.3% | 5 | 83.3% | 1 | 16.7% | USA |
| University of Oxford | 6 | 1.75% | 1 | 16.7% | 5 | 83.3% | 4 | 66.7% | 2 | 33.3% | 4 | 66.7% | 2 | 33.3% | UK |
| Yale University (*) | 6 | 1.75% | 4 | 66.7% | 1 | 16.7% | 2 | 33.3% | 3 | 50.0% | 4 | 66.7% | 2 | 33.3% | USA |
| Tel Aviv University | 5 | 1.46% | 2 | 40.0% | 2 | 40.0% | 2 | 40.0% | 2 | 40.0% | 2 | 40.0% | 3 | 60.0% | Israel |
| University of Washington | 5 | 1.46% | 3 | 60.0% | 2 | 40.0% | 2 | 40.0% | 3 | 60.0% | 3 | 60.0% | 2 | 40.0% | USA |
| Cornell University (*) | 5 | 1.46% | 2 | 40.0% | 2 | 40.0% | 1 | 20.0% | 3 | 60.0% | 3 | 60.0% | 2 | 40.0% | USA |
| Georgia Institute of Technology | 5 | 1.46% | 4 | 80.0% | 1 | 20.0% | 1 | 20.0% | 3 | 60.0% | 3 | 60.0% | 2 | 40.0% | USA |
| Rutgers University | 5 | 1.46% | 5 | 100.0% | 0 | 0.0% | 0 | 0.0% | 5 | 100.0% | 2 | 40.0% | 3 | 60.0% | USA |
| Texas A&M University (*) | 5 | 1.46% | 1 | 20.0% | 3 | 60.0% | 4 | 80.0% | 1 | 20.0% | 5 | 100.0% | 0 | 0.0% | USA |
| University of California. Davis | 5 | 1.46% | 4 | 80.0% | 1 | 20.0% | 1 | 20.0% | 4 | 80.0% | 3 | 60.0% | 2 | 40.0% | USA |
| University of Maryland | 5 | 1.46% | 2 | 40.0% | 3 | 60.0% | 3 | 60.0% | 2 | 40.0% | 4 | 80.0% | 1 | 20.0% | USA |
| Northwestern University | 4 | 1.17% | 1 | 25.0% | 3 | 75.0% | 3 | 75.0% | 1 | 25.0% | 4 | 100.0% | 0 | 0.0% | USA |
| University of California. Los Angeles | 4 | 1.17% | 2 | 50.0% | 2 | 50.0% | 2 | 50.0% | 2 | 50.0% | 3 | 75.0% | 1 | 25.0% | USA |
| University of Chicago | 4 | 1.17% | 4 | 100.0% | 0 | 0.0% | 2 | 50.0% | 2 | 50.0% | 3 | 75.0% | 1 | 25.0% | USA |
| University of Texas at Austin | 4 | 1.17% | 3 | 75.0% | 1 | 25.0% | 1 | 25.0% | 3 | 75.0% | 2 | 50.0% | 2 | 50.0% | USA |
| University of Wisconsin - Madison | 4 | 1.17% | 2 | 50.0% | 2 | 50.0% | 2 | 50.0% | 2 | 50.0% | 3 | 75.0% | 1 | 25.0% | USA |
| University of Cambridge | 4 | 1.17% | 1 | 25.0% | 3 | 75.0% | 4 | 100.0% | 0 | 0.0% | 2 | 50.0% | 2 | 50.0% | UK |